\documentstyle[12pt,aasms4,epsfig,psfig]{article}
 
\def\gta{ \lower .75ex \hbox{$\sim$} \llap{\raise .27ex \hbox{$>$}} }
\def\lta{ \lower .75ex\hbox{$\sim$} \llap{\raise .27ex \hbox{$<$}} }

\begin{document}

\title{The Jet-Disk Connection and Blazar Unification}  

\author{Laura Maraschi \& Fabrizio Tavecchio}

\affil{Osservatorio Astronomico di Brera, Via Brera 28, 20121, Milano, Italy.}

\begin{abstract}

We discuss the relation between the power carried by relativistic jets
and the nuclear power provided by accretion, for a group of blazars
including FSRQs and BL Lac objects. They are characterized by good
quality broad band X-ray data provided by the Beppo SAX satellite. The
jet powers are estimated using physical parameters determined from
uniformly modelling their spectral energy distributions (SEDs).  Our
analysis indicates that for Flat Spectrum Radio Quasars the total jet
power is of the same order as the accretion power. We suggest that blazar
jets are likely powered by energy extraction from a rapidly spinning
black hole through the magnetic field provided by the accretion
disk. FSRQs must have large BH masses ($10^8 - 10^9 M_{\odot}$) and high,
near Eddington accretion rates. For BL Lac objects the jet luminosity is
larger than the disk luminosity.  This can be understood within the same
scenario if BL Lac objects have masses similar to FSRQ but accrete at
largely subcritical rates, whereby the accretion disk radiates
inefficiently.  Thus the ``unification'' of the two classes into a single
blazar population, previously proposed on the basis of a spectral
sequence governed by luminosity, finds a physical basis.

\end{abstract}

\section{Introduction}

The formation of highly relativistic jets in Active Galactic Nuclei is
one of the most fundamental open problems in astrophysics. It is
currently assumed that jets are produced close to the central black hole,
involving power extraction from the black hole spin (Blandford \& Znajek
1977) and/or from the accretion disk (Blandford \& Payne 1982).
In both scenarios the magnetic field must play a major role in
channelling power from the BH or from the disk into the jet; in both
cases it should be sustained by matter accreting onto the BH, leading to
expect a relation between the accretion power and the jet power.

Estimates of the power of jets and of the associated accretion flows can
therefore be crucial to shed light on the jet-disk connection .  In a
pioneering work Rawling \& Saunders (1991) addressed this question
studying a large sample of radio galaxies.  They used the narrow line
luminosity as indicative of the accretion power and estimated the power
transported by the jet from the energy content and lifetime of the radio
lobes, finding a good correlation between the two. This result has been
confirmed with larger and deeper samples and with different power
``estimators'' (e.g. Willot et al. 1999; Xu, Livio \& Baum
1999). However, due to the indirect nature of the estimators used, the
``calibration'' of the relation in terms of jet power and accretion power
remains uncertain.  Celotti, Padovani \& Ghisellini (1997, hereafter CPG)
first investigated the relation between jet and disk using the direct
radio emission of the jet close to the nucleus, as resolved by VLBI, for
the jet power estimation. They considered a large sample of objects (55),
mostly blazars (including 12 BL Lac objects) and derived the accretion
luminosity from the broad emission lines when available. They found a
suggestive hint of correlation between these two quantities, although the
statistical significance was too low to draw a firm conclusion.

Blazars are in fact the best laboratories to study the physics of
relativistic jets. Their emission (from radio to gamma-rays) is dominated
by the beamed non-thermal continuum produced in the jet (Urry \& Padovani
1995).  Their SEDs are in general well understood as synchrotron plus IC
emission. When observations of both components are available, the basic
physical quantities of the emission region can be derived in a robust
way, allowing to estimate the jet power in the region closest to its
origin (e.g. Tavecchio et al. 2000, hereafter Paper I).  Particularly
interesting for the study of the jet-disk connection are those blazars
showing thermal features directly related to the central accretion flow,
such as the so-called blue bump and/or the bright emission lines produced
in the Broad Line Region.

In the present work we consider a relatively small group of bright
blazars (11 FSRQs and 5 BL Lacs) with good wavelength coverage, basing
jet power estimates on physical parameters derived from uniformly
modelling the observed SEDs (Tavecchio et al. 2002, hereafter Paper
II). All these objects have broad band X-ray spectra (0.1 - 100 keV) from
{\it Beppo}SAX observations. As discussed below (section 2) the
information on the X-ray continuum is crucial in order to have reliable
estimates.

The plan of the paper is as follows. In Section 2 we describe the objects
and the adopted method for estimating the power of the jet.  We then
compare the jet powers with their luminosities (integrated over angles),
deriving the jet radiative efficiencies, and the jet luminosities with
the accretion luminosities estimated from the blue bump or line strengths
(Section 3). The derived relation between jet luminosity and accretion
luminosity is discussed in Section 4 taking into account the radiative
efficiencies and compared with the expectation of current models.
Implications for a scenario of blazar unification are considered.  The
conclusions are presented in Section 5.  Throughout the paper we assume
$H_0=50$ km s$^{-1}$ Mpc$^{-1}$ and $q_0=0$.

Preliminary results were presented in Maraschi \& Tavecchio (2001a)
Maraschi \& Tavecchio (2001b) and Maraschi (2001).

\section{ Sample and General Model}

In view of a reliable estimate of the power carried by the relativistic
jet we require uniform "good quality" data on their SEDs.  These are
commonly modelled with synchrotron and IC radiation from a uniform
emission region. Both components must be observationally constrained in
order to derive the physical parameters of the emitting region (Tavecchio 
et al. (1998); see further discussion below).

In practice the broad band coverage (0.1-100 keV) allowed by the BeppoSAX
satellite is essential.  In fact, in the case of FSRQs the X-ray emission
derives from the Inverse Compton mechanism (with seed photons external to
the jet (EC) and in some cases an additional contribution from the
synchrotron photons (SSC)).  The X-ray data then fall on the low energy
branch of the IC component, constraining the electron spectrum at low
energies, which is particularly important in the power estimate (see
below).  In the case of BL Lacs, whose SEDs peak at higher frequencies,
the X-ray data constrain the position of the synchrotron peak, while
information on the IC peak is provided by observations in the $\gamma
$-ray (GeV range) or in the TeV domain.

Specifically the selection criteria for the sources considered are the
 following. The FSRQs belong to a subsample of the 2-Jy catalogue by
 Padovani \& Urry (1992; 50 sources) with a threshold of $F_{\rm 1\,
 keV}>0.5 \mu$Jy (19 objects) chosen in order to obtain good {\it
 Beppo}SAX spectra up to 100 keV. 12 objects have been observed until
 now.  In Paper I we discussed three of them, namely 0836+710,
 1510-089 and 2230+114; another 6 sources are analyzed in Paper II
 (0208-512, 0521-365, 1641+399, 2223-052, 2243-123, 2251+158) and 2
 sources (3C279, PKS 0537-441) are discussed elsewhere (Maraschi et
 al. 1998; Ballo et al. 2002; Pian et al., submitted). For 0528+134, for
 which good data on the SED are available (Ghisellini et al. 1999) we
 were unable to find the line luminosity in the literature and for this
 reason it was excluded from our sample.

The 5 BL Lac objects were chosen so as to have a well measured SED near
the synchrotron peak complemented by information on the IC peak.  They
are the two TeV sources Mkn 421 and Mkn 501 (for which simultaneous
observations are used; Maraschi et al. 1999, Tavecchio et al. 2001), PKS
2155-304 (Chiappetti et al. 1999), ON231 (Tagliaferri et al 2000) and BL
Lac itself (Ravasio et al. 2002). Other BL Lacs with $Beppo$SAX data and
$\gamma $-ray information (from EGRET) are 0716+714 and 0235+164, but for
0716+714 the redshift is unknown, while 0235+164 shows a peculiar
phenomenology, possibly due to gravitational lensing (e.g. Webb et
al. 2000). Both were therefore excluded from the present analysis.

The observational information on the SEDs is given in the above papers.
The physical parameters of the jets were rederived uniformly reproducing
all the SEDs with a synchrotron + inverse Compton model including both the
synchrotron and external photons as seed photons (SSC+EC). 

We adopt a one-zone model, in which the radiation is produced in a
homogeneous emitting region by a single electron population.  One-zone
models are supported by a number of observational evidences, at least for
the spectral range from the gamma-ray band down to the optical-IR
band. In particular, the observations of correlated variability at
different frequencies suggest co-spatial production of low and high
energy photons via the two mechanisms (synchrotron + IC) by the high
energy branch of the electron population (e.g. Ulrich, Maraschi \& Urry
1997; Urry 1999; Giommi et al. 1999, Maraschi et al. 1999; Sambruna 2000;
Sikora \& Madejski 2001). In general the one-zone model predicts
synchrotron self absorption in the FIR /submm range. Thus it cannot
account for the radio emission, which is thought to be due to the
superposition of self-absorbed synchrotron components produced further out
($d \gta 0.1$ pc) along the jet (e.g. Blandford \& Konigl 1979). The one
zone model considered here naturally appears as the innermost component
of the inhomogeneous jet model.
 
As shown in Tavecchio et al. (1998) for the case of a one zone
synchrotron+SSC model, the knowledge of the frequency and luminosity of
both the synchrotron and IC peaks, together with the upper limit on the
size of the emission region derived from variability, allows to
univocally determine the model parameters.  In the case, particularly
relevant for the present work, in which the IC component is dominated by
scattering of external radiation, the model involves an additional
parameter, that is the energy density of the external radiation field.
However (see e.g. Paper I and the discussion below) the latter quantity
can be estimated with reasonable confidence from the available
luminosities of the emission lines and/or the UV bump. Therefore, given
sufficient observational information on both peaks, in both the SSC and
EC models reliable estimates of the basic physical quantities of the jet
can be derived.

For a complete description of the model we refer to the Appendix.
We just recall that a phenomenological description of the electron
spectrum was adopted (analogous to a broken power-law):
\begin{equation} 
N(\gamma )=K\gamma ^{-n_1}\left( 1+ \frac{\gamma }{\gamma _b
}\right)^{n_1-n_2} \,\,\,\,\,\,\,\,\,  \gamma _{\rm min} < \gamma < 
\gamma _{\rm max}
\end{equation} 
\noindent 
where $K$ is the normalization factor, $\gamma _b$ is the Lorentz factor
of electrons at the spectral break, $n_1$ and $n_2$ are the spectral
indices below and above the break, respectively and $\gamma _{\rm min}$
and $\gamma _{\rm max}$ are the minimum and the maximum energies of the
relativistic electrons. For Mkn 501 we used a slighly modified
electron energy distribution (see Tavecchio et al. 2001), with the
addition of an exponential high energy cut-off.

The viewing angle is assumed to be $\theta \sim 1/\Gamma $ where $\Gamma
$ is the bulk Lorentz factor of the emitting plasma; in these conditions
the Doppler factor $\delta \simeq \Gamma$. This choice, which eliminates
the angle to the line of sight as an independent parameter, may appear
arbitrary for individual objects; however, it can be justified for a group
of objects. In fact, for fixed $\delta $ (derived from the spectral
modeling), the probability of observing a source at an angle $\theta $ is
maximal around $\theta \sim 1/\Gamma $, since this is the maximum angle
allowed for a given $\delta $: thus, for a group of sources, as in the
case discussed here, we expect that the {\it average} viewing angle is
close to $1/\Gamma $. A known exception is the case of 0521-365, for
which various indicators suggest that the jet forms a relatively large
angle with the line of sight (Pian et al. 1996).  Assuming that the bulk
Lorentz factor of the emitting plasma is similar to that of the other
blazars, $\Gamma _{\rm b} \sim 10$, this implies that the emission from
0521-365 is weakly boosted. Here we assumed $\Gamma=10$ and $\theta
=15^{\rm o }$, implying $\delta=3$.

The external radiation field is described as a black-body with
temperature $T\simeq 10^4$ K and energy density $U_{\rm ext}$.
Uncertainties in the temperature (to within factors of a few) do not
strongly affect the determination of other physical parameters. The most
important parameter is the energy density $U_{\rm ext}$. The latter can
be derived from the luminosity of the broad line region (BLR) and/or of
the accretion disk by $U_{\rm ext} = L_{\rm BLR}/4\pi R_{\rm BLR}^2c$ and
$L_{\rm BLR}=\tau L_{\rm disk}$ (where $\tau $, usually assumed to be
$\sim 0.1$ represents the fraction of the central emission reprocessed by
the BLR and $R_{\rm BLR}$ its radius ). For sources showing a clear UV
bump we used $L_{\rm disk}=L_{\rm UV}$, while for the other cases $L_{\rm
BLR}$ was derived from the luminosity of the observed broad emission
lines, applying correction factors as used by CPG. $R_{\rm BLR}$ was
adjusted in the fits but the derived values were found to agree within a
factor of 2 with those predicted by the correlation of Kaspi et
al. (2000) between $R_{\rm BLR}$ and $L_{\rm BLR}$.

In the case of BL Lac, ON231, 0836+710, 1510-089 and 2230+114 the models
for the SEDs presented in the original papers (Ravasio et al. 2002,
Tagliaferri et al 2000, Paper I) were computed with different hypothesis
on the electron energy distribution (Ghisellini et al. 1998).  Electrons
were assumed to be continuously injected in the emitting region, with a
power-law distribution with index $n_{\rm inj}$, extending from $\gamma
_{\rm min}$ to $\gamma _{\rm max}$. The equilibrium distribution reached
as a result of cooling has then a double-power law shape, with energy
indices $n_1=2$ for $\gamma < \gamma _{\rm min}$ and $n_2=n_{\rm inj}+1$
for $\gamma > \gamma _{\rm min}$. Therefore, in this model the spectral
index of the low-energy portion of the emitted spectrum is fixed to be
$\alpha _1=0.5$ and the peak of the SED corresponds to electrons with
$\gamma = \gamma _{\rm min}$, which plays the same role as $\gamma _b$ in
the model adopted in the present paper. Notably, although some of the
parameters obtained with the two versions of the model are different,
important derived quantities such as powers and luminosities (see next
sections) appear to be rather stable (within a factor of 2-3) against the
details of the model adopted, supporting the consistency of our
results. Table 1 reports the list of sources, their redshifts and the
values of the parameters derived from modeling the SEDs.

\section{Radiative Jet Luminosity and Power}

The kinetic power of the jet, i.e. the energy flux of the relativistic
flow through a section $\pi R^2$ of the jet, is given by: $P_{\rm
jet}=\pi R^2 \beta c \,U \Gamma ^2$ (e.g, CPG) where $U=U_B+U_e+U_p$ is
the total energy density in the jet frame, due to magnetic field,
relativistic electrons and, if present, protons ($L_{\rm jet}$ = $ L_{\rm
B} + L_{\rm e} + L_{\rm p}$).  The energy density in particles is given
by $U_e + U_p=n_em_ec^2[<\gamma >+(n_p/n_e) (m_p/m_e)]$ where
$n_e=\int_{\gamma _{\rm min}} ^{\gamma _{\rm max}} N(\gamma)d\gamma $ is
the electron density and $<\gamma >$ is the average Lorentz factor of
electrons.
 
The critical parameter determining $P_{\rm jet}$ is the total number of
particles, which in turn depends on the energy spectrum of the electrons
below $\gamma_b$ and on the value of $\gamma _{\rm min}$.  For FSRQ both
quantities can be inferred from the shape of the X-ray spectrum.
In particular values of $\gamma _{\rm min} >> 1$ would produce an
unobserved break in the X-ray continuum (e.g. Paper I). Moreover in several
cases our data exclude an important contribution from cold pairs, whose
presence should produce a ``bump'' in the soft X-ray spectrum (Sikora et
al. 1997). 

We will assume that the jet is composed by a normal plasma with 1 (cold)
proton per relativistic particle. This hypothesis is justified below.
For uniformity we evaluate $P_{\rm jet}$ with $\gamma _{\rm min}=1$ for
all the objects: the power should then be considered as an {\it upper
limit} to the actual power of the jet. An increase of a factor 10 in
$\gamma_{\rm min}$ (which cannot be excluded in all cases) would lower
the estimated powers by a factor 5-10, depending on the value of
$n_1$. 

For BL Lac objects X-rays can derive from the synchrotron component or
from IC emission produced through SSC. In both cases the emission is
produced by high-energy electrons, $\gamma >1000 $ in the case of
synchrotron and $\gamma \sim 100$ for SSC. Thus the SEDs contain less
stringent information on the amount of low energy particles in the jet,
in particular on the value of $\gamma _{\rm min}$ and on the index
$n_1$. Since we work in the perspective of a unified model we will assume
also for BL Lacs an equal number of protons and electrons and $\gamma
_{\rm min}\sim 1$.

The derived jet powers are reported in Table 2.  Table 2 shows that the
magnetic field tends to be close to equipartition with the relativistic
particles in FSRQs but largely below equipartition in BL Lacs,
especially in TeV sources ($U_e/U_B\sim 10-100$). The latter result has
been recently independently found by other authors and appears to be
rather robust (see in particular the discussion of Kino et
al. 2001). Clearly in all cases the total jet power is dominated by
protons, while the magnetic field and relativistic electrons give minor
contributions.

An important quantity is the total radiative luminosity ($\L_{\rm jet}$)
of the jet, integrated over the whole solid angle in the observer frame.
This is derived from the observed apparent luminosity correcting for
beaming (e.g., Sikora et al. 1997): $L_{\rm jet}= \frac {L_{\rm obs}}
{\delta^4} \Gamma ^2 \simeq \frac {L_{\rm obs}} {\Gamma ^2}$.  $L_{\rm
jet}$ represents the {\it minimum} power that must be associated with the
jet in order to produce the observed luminosity, i.e. a {\it lower limit}
to $P_{\rm jet}$.

In Fig 1 we compare the radiative luminosities of jets with the powers
provided by the electron component only. The dotted line indicates the
relation $L_{\rm e}=L_{\rm jet}$. In most cases the power associated with
the electron component alone is is insufficient or at best marginally
sufficient to sustain the jet beyond the inner emission region. 
Analogously, the Poynting flux associated with the transport of 
magnetic field is too small, unless other components or complex
geometries for the magnetic field are invoked. Thus a proton
contribution seems the most natural to explain the transport of energy to large
distances.

In Fig. 2 $L_{\rm jet}$ is compared with the total power estimated
including protons $P_{\rm jet}$. There is a well defined correlation
between the two quantities (probability$>$99.9\%, slope=$1.12\pm 0.17$),
extending over a range of about 4 orders of magnitude. Notably, BL Lac
objects appear to lie on the same (linear) correlation with powerful
quasars, supporting the view that these two classes of Blazars have
similar jets.

The radiative efficiency, $\eta = L_{\rm jet}/P_{\rm jet}$, turns out to
be in the range 10 -- 1 \%. It is interesting to note that similar values
are naturally predicted by the internal shock scenario, recently proposed
for Blazars by Spada et al. (2001).

\section{The Jet-Disk Connection}

The disk luminosity $L_{\rm disk}$ was estimated either directly from the
optical-UV luminosity of the blue bump attributed to an optically thick
accretion disk (e.g. Sun \& Malkan 1989) or from the luminosity of the
broad emission lines (assuming $\tau =0.1$), using the relations proposed
by CPG. For 3C279 we used the luminosity of the blue bump identified in
the IUE data by Pian et al (1999) we checked that this luminosity is
close to 10 times the luminosity estimated from the emission lines.  For
BL Lacs, except for BL Lac itself, for which a broad emission line has
been observed (Corbett et al. 2000) one can only derive upper limits to
the luminosity of the (putative) accretion disk (CPG).

Fig. 3 shows the radiative luminosity of the jet, $\L_{\rm jet}$, against
 the disk luminosity $L_{\rm disk}$. A dotted line represents the
 equality of the two. It is apparent that for high luminosity blazars
 (FSRQ) $\L_{\rm jet}$, which represents the minimal power transported by
 the jet, is {\it of the same order} as the luminosity released in the
 accretion disk.  The situation is different for low luminosity blazars
 (BL Lac objects). For the latter objects the jet luminosity is {\it
 higher } than the estimate/upper limits on their disk luminosity.  Since
 $\L_{\rm jet}$ is obtained from the observed SED with only a beaming
 correction the results above are largely independent of the theoretical
 model adopted and in particular of the assumptions concerning the proton
 component which enter in the estimate of $P_{\rm jet}$.

In view of a more quantitative discussion it is however essential to
convert the derived luminosities into powers taking into account
radiative efficiencies for both the jet and the disk. Let us define
$L_{\rm jet}=\eta P_{\rm jet}$ and $L_{\rm disk}=\epsilon P_{\rm
acc}$. As discussed in Sect. 3, $\eta \sim 0.1-0.01$, where the higher
value holds for high luminosity jets. On the other hand, for high
luminosity blazars where the blue bump and/or broad lines are observed,
the accretion disk is also extremely luminous (see Fig. 3). It is then
natural to assume that the disk should have efficiency close to standard,
that is $\epsilon \simeq 0.1$, otherwise implausibly high accretion
rates, as large as 10 solar masses per year would be required.  The
radiative efficiencies for the jet and disk are then of the same order so
that comparable luminosities $L_{\rm jet} \simeq L_{\rm disk}$ imply
comparable powers : $P_{\rm jet} \simeq P_{\rm acc}$.  This near
equality, though difficult to achieve according to presently available
models (see below), represents the main result of our analysis.

For low luminosity blazars $\eta \sim 0.01$ and $\L_{\rm jet} > L_{\rm
disk} $. Assuming that the relation $P_{\rm jet} \simeq P_{\rm acc}$ is
verified for all blazars, in order to account for a jet luminosity {\it
larger} than the disk luminosity a very low radiative efficiency for the
disk is implied.  This can be naturally explained if the accretion flow
in these systems has the structure of an "ion torus" (Rees et al. 1982)
or ADAF (Narayan et al. 1998) whereby the inefficient cooling keeps the
flow geometrically thick, supported by the pressure of hot ions. Such
configurations are possible if the accretion rate is largely
sub-Eddington.

This scenario then suggests that the range of powers observed in
blazars is due essentially to a range in accretion rate onto black
holes of equally large masses.

\subsection{Discussion}

Two main classes of models for the formation of jets consider either
extraction of rotational energy from a rapidly spinning black hole, the
Blandford \& Znajek (1977, BZ) process, or magnetohydrodynamic winds
arising from the inner regions of accretion disks (MHD) (Blandford \&
Payne 1982). In the latter scenario jets would be powered solely from the
accretion process through the action of the magnetic field. Livio et
al. (1999) show that the power extracted through the latter mechanism can
be important. For reasons of global consistency it would seem however
difficult that in this type of model a large fraction (of order 1) of the
accretion power could be channeled into a highly relativistic outflow.

The complex analysis of BZ can be summarized in the well known expression:
\begin{equation}
P_{BZ}\simeq \frac{1}{128}B_0^2 r_g^2 a^2 c  \,\,\,\,\,
\end{equation}
where $r_g$ is the gravitational radius and $a=j/j_{\rm max}$ the
adimensional angular momentum of the BH, $a=1$ for
a maximally rotating black hole (e.g. Thorne et al. 1986). The
critical problem is the estimate of the intensity reached by the magnetic
field threading the event horizon, which must be provided by the
surrounding matter.

Let us take as reference an extreme approximation i.e.  a spherical free
fall accretion flow.  Assuming equipartition of magnetic and kinetic
energy density ($B_0^2/ 8\pi \simeq \rho c^2$) as a 0 order approximation
to the physical picture envisaged for the "plunging" region (Krolik
1999), it is easy to find that

\begin{equation}
P_{BZ} = g a^2 P_{acc}= g a^2 \dot{m} 10^{47} M_9 a^2 \,\,\,\,\, {\rm erg/s}
\end{equation}
where $P_{acc}=\dot{M}c^2$ is the accretion power and $g$ is
$1/64$. $\dot{m}=\dot{M}c^2/L_{\rm Edd}$ is the accretion rate in
Eddington units and $M_9$ is the black hole mass in units of $10^9$
$M_{\odot }$.  This simple formula shows clearly that even when the jet
is produced at the expense of the black hole rotational energy the
generated power is closely linked to the accretion rate.
Eq. (3) is shown as a dashed line in Fig.3 (assuming
$\eta\simeq\epsilon\simeq 10^{-1}$) and is clearly insufficient 
to account for our results.

In the case of disk accretion one expects higher densities and plausibly
 higher fields. Ghosh \& Abramovicz (1997; GA) discussed the possible
 field strengths threading the black hole horizon on the basis of the
 disk model of Shakura and Sunyaev (1973). Their results are shown in
 Fig.3 as continuous lines (efficiencies as above).  It is interesting to
 note that when the disk is in the gas pressure dominated regime
 ($\dot{m} < 10^{-3}$) ratios $L_{\rm jet} / L_{\rm disk}$ not far from
 unity are indeed obtained ($g\simeq 1$).  However, due to the pressure
 saturation introduced by the formation in the disk of a radiation
 pressure dominated region (horizontal branches in Fig.3), the model
 fails to explain the large powers observed in the jets of bright
 quasars, even for maximal rotation and large BH masses ($g<<1$ for
 $\dot{m} > 10^{-3}$).

Our results, pointing to a high yield for the jet production mechanism,
underline the need of further investigations of the BZ process in more
general conditions, both for the disk model (e.g. $\beta $-disk models
Sakimoto \& Coroniti 1981) and for its interaction with a fast spinning
black hole.  In fact it has been suggested that dynamical effects and
frame dragging by the rotating hole may restore $g$ to values of order 1
or even larger (Krolik 1999, Meier 1999, 2001).

The scenario indicated by our results involves a substantial equality
of the jet and accretion powers which could hold for all blazars.
High luminosity blazars including highly polarised, optically
violently variable quasars (HPQ, OVV) or more generically
quasars with flat spectrum radio cores (FSRQ)  owe their properties
to a high, near critical accretion rate, which accounts at the same
time for the presence of bright accretion disks and of powerful jets.
Low luminosity blazars (otherwise called BL Lacs) where clear
signatures  of an accretion disk are not found, 
can be explained by  largely subcritical accretion rates,
giving rise to radiatively inefficient accretion flows and low
power jets. Thus there is no "genetic" difference between 
FSRQs and BL Lac objects and the blazar population can be "unified"
and described in terms of a single parameter ($\dot m$).

The idea that the power scale of blazars corresponds to a scale of
accretion rates in objects with essentially similar (large) masses and
high angular momentum ($a\sim 1$) also provides a physical basis to
understand the "spectral sequence" of blazars proposed by Fossati et al.
(1998). The latter paper showed that the spectral energy distributions
(SED) of blazars change systematically with luminosity in the sense of a
shift of the emission peaks towards higher frequencies with decreasing
luminosity. The modelling (Ghisellini et al. 1998) showed that the
particles radiating at the peaks have lower energies in higher luminosity
objects, which was interpreted as due to a larger density of ambient
photons resulting in a larger cooling rate.  The scaling proposed here
supports this view, in the sense that, due to the ADAF like accretion
flow, less powerful jets find a much cleaner ambient in the transition
region from the black hole to the parsec scale.

The evolutionary aspects of this scenario have been explored by Cavaliere
\& D'Elia (2002) and found to be in agreement with present data on the
number counts.  Moreover Cavaliere \& D'Elia (2002) also discuss reasons
for the correlation between luminosity and the average shapes of the
SEDs. Within a closely similar scenario B\"ottcher \& Dermer (2002)
developed a specific model of the expected SEDs introducing a
number of hypotheses and parameters.

In particular Cavaliere \& D'Elia (2002) and B\"ottcher \& Dermer (2002)
proposed that the blazar spectral sequence also traces an evolutionary
sequence, from young FSRQs to the older BL Lac objects. FSRQs are rich of
gas and therefore are characterized by large accretion rates while BL Lac
objects represent evolved sources depleted of gas, with faint nuclear
emission and low power jets. Our results provide a solid basis to these
speculations.

This scenario can be observationally tested taking advantage of the
correlations between the mass of the central black hole and the host
galaxy properties (e.g. Merritt \& Ferrarese 2002) for estimating the
black hole masses in blazars of different types.  Treves et al. (2001)
measured the host galaxies around a large number of BL Lacs. Their
results (see also Urry et al. 2000; Scarpa et al. 2000) show that the
magnitudes of the host galaxies have relatively little scatter and are
independent of the luminosity of the BL Lac. Assuming (e.g. Ferrarese \&
Merritt 2000) that the central black hole mass correlates with the mass of
the bulge and therefore with the magnitude of the galaxy one can derive
(to zero order) that the BH mass is similar in all these objects. More
accurate estimates are possible measuring the stellar velocity
dispersions of the host galaxies for which the intrinsic correlation with
black hole mass is thought to be much tighter. Using the latter method
Barth, Ho \& Sargent (2002) derived a black hole mass of $10^9$
$M_{\odot}$ for Mkn 501 yielding for this object a highly subcritical
accretion rate, in complete agreement with our expectations.

\section{Conclusions}

Estimates of the powers transported by the jets of a small group of FSRQs
and BL Lac objects, for which broad band spectra have been obtained with
{\it Beppo}SAX, were derived by modelling their overall SEDs with the
widely accepted Synchrotron/Inverse Compton emission model. Comparing jet
powers and luminosities with estimates of the accretion luminosity
derived from the optical-UV spectra, we find that for the most powerful
blazars the power carried by the jet is of the same order as the
accretion power.  Moreover comparing the respective jet and accretion
luminosities of FSRQ with those of BL Lacs with data of comparable
quality on their SEDs, we find that for the latter the jet luminosity is
higher than the upper limits on the accretion luminosity.

Taking into account the radiative efficiencies of both the jet and the
accretion disk, we infer that the mechanism of jet production must have
high efficiency (in terms of $\dot{M}c^2$) favoring energy extraction
from a Kerr Hole rather than a hydromagnetic wind generated by an
accretion disk.

In view of the various approximations used the conclusions above can only
be regarded as tentative. Nevertheless the available evidence suggests
that the main parameter governing the total power and the ratio between
jet and accretion luminosity is the {\it accretion rate}.  In FSRQs the
accretion rate must be high, near Eddington. This can explain the large
powers and the contemporaneous presence of thermal signatures associated
with efficient disk accretion. The absence of thermal signatures in BL
Lac objects (and in low power radio galaxies) can be ascribed to a highly
subcritical accretion disk with low radiative efficiency.

The unification of the two classes of sources into a single blazar
population, previously proposed on the basis of a spectral sequence
governed by luminosity (Fossati et al.,1998; Ghisellini et al.,1998) and
recently revisited by Cavaliere \& D'Elia (2002) and B\"ottcher \& Dermer
(2002) finds therefore a physical basis.

The scenario is testable through measurements of the properties of the
host galaxies of blazars of different types, leading to estimates of the
central black hole mass.

\acknowledgments{This work was partly supported by the Italian Ministry
for University and Research (MURST) under grant Cofin98-02-32 and by the
Italian Space Agency (I/R/037/01).}

\appendix
\section{The homogeneous synchrotron-Inverse Compton model}

We give here the full description of the model used to
reproduce the SED of blazars adopted in our work.

The emission region is assumed to be a sphere (``blob'') with radius $R$,
uniformly filled by a tangled magnetic field with intensity $B$ and
isotropic relativistic electrons with energy distribution $N(\gamma
)$. The region is in motion with velocity $\beta c$ and bulk Lorentz
factor $\Gamma $ at an angle $\theta $ with respect to the line of
sight. Relativistic effects in the emitted radiation are then taken into
account by the relativistic Doppler factor $\delta $, defined by:
\begin{equation} 
\delta =\frac{1}{\Gamma(1-\beta \cos \theta)}
\end{equation} 
\noindent
The electron distribution is described (for $\gamma _{\rm min}< \gamma
< \gamma _{\rm max}$) by the law (in the following physical quantities
are expressed in the comoving frame):
\begin{equation} 
N(\gamma )=K\gamma ^{-n_1}\left( 1+ \frac{\gamma }{\gamma _b
}\right)^{n_1-n_2} 
\end{equation} 
\noindent 
where $K$ [cm$^{-3}$] is a normalization factor (it represents the
density of electrons with $\gamma =1$), $\gamma _b$ is the break Lorentz
factor, $n_1$ and $n_2$ are the spectral indices below and above the
break, respectively. This law represents a double power-law distribution
with a smooth connection. This particular form for the distribution
function has been assumed on a purely phenomenological basis, in order to
describe the curved shape of the SED.

Once the parameters are specified, the outcoming spectrum is calculated
using the standard single-electron synchrotron emissivity (e.g. Rybicki
\& Lightman 1977) and the IC emissivity including the full Klein-Nishina
cross section given by Jones (1968). Specifically:

\noindent
- In the case of the synchrotron emission the emissivity at a given
frequency $\nu _s $ is calculated using the relation:
\begin{equation}
j_s(\nu _s  )=\frac{1}{4\pi }\int _{\gamma _{\rm min}}^{\gamma _{\rm
max}} N(\gamma) P(\nu _s  , \gamma) d\gamma
\label{synchro}
\end{equation}
\noindent
where $P(\nu _s , \gamma)$ is the standard specific power emitted by a
single electron with Lorentz factor $\gamma $. The spectrum is calculated
between the two limit frequencies $\nu _{s,1} $ and $\nu _{s,2} $, where
$\nu _{s,1} $ is the self absorption frequency (calculated using the
approximation (obtained for the slab geometry) given in Ghisellini et
al. 1985) and $\nu _{s,2} $ is the maximum frequency, evaluated as the
typical synchrotron frequency of electrons with energy $\gamma _{\rm
max}.$, $\nu _{s,2} \sim 2.8 \times 10^{6} B \gamma _{\rm max}^2$.

\noindent
- For the calculation of the IC spectrum we adopt the single-electron
emissivity $j_C(\nu _C ; \gamma, \nu _{\rm t})$ (function of the electron
energy $\gamma mc^2$ and of the soft photon frequency $\nu _{\rm t}$)
calculated by Jones (1968) (see also Blumenthal \& Gould 1970). The
total emissivity $j_C(\nu _C)$ is calculated by integrating the
single-electron emissivity over the soft photon spectrum and the electron
energy distribution (from $\gamma _{\rm min}$ to $\gamma _{\rm max}$):
\begin{equation}
j_C(\nu _C )= \int _{\gamma _{\rm min}}^{\gamma _{\rm
max}} N(\gamma)\int _{\nu _{\rm t, min} }^{\nu _{\rm t,
max} }n_t (\nu _{\rm t} ) j_C(\nu _C; \gamma, \nu _{\rm t} ) d\nu _{\rm
t}  d\gamma 
\label{IC}
\end{equation}
\noindent
where $n_t (\nu _{\rm t} )$ is the numerical density of
target photons, $n_t (\nu _{\rm t} ) = U (\nu
_{\rm t} )/h\nu _{\rm t} $. 
The energy density $U (\nu _{\rm t} )$ of the soft target
photons is calculated as follows:

\noindent
$\bullet $ In the case of \underline{SSC emission} the energy density of the
synchrotron target photons is calculated with (e.g. Ghisellini et al. 1998):
\begin{equation}
U (\nu )=\frac{4\pi R}{3c} j_s (\nu
 )
\label{uradsinc}
\end{equation}
\noindent
where $\nu _{\rm t, min} $ and $\nu _{\rm t, max} $ are
fixed to $\nu _{s,1} $ and $\nu _{s,2} $, respectively.

\noindent
$\bullet $ For the calculation of the \underline{EC spectrum} we need a
prescription to model the external radiation field. We assume that the
spectrum of the external radiation is characterized by a blackbody-like
spectrum with temperature $T$ and total luminosity $L_{\rm ext}$, diluted
in a spherical region with radius $R_{BLR}$ (typically of the order of
the radius of the Broad Line Region). The external radiation energy
density (in the \underline{observer frame}) is:
\begin{equation}
U_{\rm obs}^{\rm ext}(\nu _{\rm obs})=\frac{L_{\rm ext}(\nu _{\rm obs})}{4\pi
R_{BLR}^2 c}
\label{uext}
\end{equation}
\noindent
Due to relativistic amplification effects the (angle averaged) energy
density seen in the blob's reference frame will be:
\begin{equation}
U ^{\rm ext}(\nu )=\Gamma U_{\rm obs}^{\rm ext}(\nu _{\rm obs}=\nu / \Gamma). 
\label{uextcom}
\end{equation}
\\

Comoving emissivities are used to calculate the observed fluxes as follows:

\noindent
-Synchrotron and SSC emissions are transformed according to the standard
relations (e.g. Lind \& Blandford 1985):
\begin{equation} 
F_{\rm obs}(\nu _{\rm obs}) =  \frac{\delta ^3}{D^2} j(\nu =\nu _{\rm
obs}/\delta) V  
\label{fampl}
\end{equation}
\noindent
where $D$ is the luminosity distance and $V$ the comoving emitting
volume.

\noindent
-As pointed out by Dermer (1995) the beaming of the external radiation
 field in the source frame introduces a supplementary $\delta $ term in
 the calculation of the final flux. Dermer's calculation assumed a number
 of approximations, namely a single power-law electron distribution, the
 Thomson cross-section, a monochromatic external radiation field and
 extreme beaming (assuming that photons enter the source only head-on):
 thus we can not directly apply his results (expressed by the
 supplementary term $\delta ^{1+\alpha}$, where $\alpha $ is the
 power-law index of the IC spectrum) to our more complex model. However,
 in the case $\delta = \Gamma$ (as assumed in the present work), the
 effect of the anisotropy is taken into account simply assuming
 Eq.(\ref{fampl}) and using Eq.(\ref{uextcom}) for the energy density.

\clearpage

\vskip 1.5 truecm

\centerline{ \bf Figure Captions}

\vskip 1 truecm

\figcaption[pj]{Radiative luminosity vs. the power transported by
relativistic electrons. The dashed line indicates the relation $L_{\rm
jet}=P_e$.\label{pj}}

\figcaption[rj]{Radiative luminosity vs. jet power for the sample of
Blazars discussed in the text (open circles represent BL Lac
objects). The dashed line indicates the least-squares fit to the
data. \label{rj}}

\figcaption[dj]{Radiative luminosity of jets vs disk luminosity. The
dotted line indicate the relation $L_{\rm jet}=L_{\rm disk}$. The solid
lines represent the {\it maximum} jet power estimated for the Blandford
\& Znajek model for black holes with different masses (in Solar
units). The dashed line is obtained with a simple spherical free fall
approximation. \label{dj}.}

\begin{table*}
\begin{center}
\caption{Parameters used for the emission model described in the text
(see text for definitions).}
\begin{tabular}{llcccccccc}
\\ 
\hline 
\hline 
Source & $z$ &$R$ & B & $\delta$ & $\gamma _{\rm b}$ & n$_1$& $n_2$ & $K$ & $U_{\rm BLR}$ \\
& & $10^{16}$ cm& G& & & & & cm$^{-3}$ & $10^{-3}$ erg cm$^{-3}$\\ 
\hline

0208-512(PKS)& 1.003 &1.5& 1.5 & 18& 100& 1.4& 3.8& $2\times 10^4$& 15\\

0521-365(PKS)& 0.055&2.0& 0.3& 3$^*$ & $8.8\times 10^3$  &  1.25& 4 &$3\times 10^3$
& 0.1\\

0537-441(PKS)& 0.89 & 4.65 & 2.1 & 10 & 400 & 1.6 & 3.4& $3.5\times 10^3$ &
33\\  

0836+710(4C71.07) & 2.172 &4& 3& 16& 50& 1.6& 4.0& $5\times 10^4$& 54\\

1253-055(3C279)& 0.538 & 5 & 0.5 & 12.3 & 600 & 1.6 & 4.2 & $4.5\times 10^3$ &
0.1\\

1510-089(PKS)& 0.361& 3& 1.5& 19& 50& 1.7& 3.6& $6\times 10^3$& 0.8 \\

1641+399(3C243)& 0.593& 4.0& 2.9& 9.75& 200& 1.5& 4.2& $2.8\times 10^3$&
30 \\

2223-052(3C446)& 1.4&4.25 & 5.6& 17& 135& 1.6& 4.3& $1.7\times 10^3$& 18\\

2230-014(CTA102)& 1.037& 3& 1.65& 18& 55& 1.9& 3.4& $3\times 10^4$& 6.5\\

2243-123(PKS)& 0.63& 3.5 & 2.5& 15& 250& 1.6& 4.3& $1.7\times 10^3$& 18 \\

2251+158(3C454.4)& 0.859& 4.0 & 1.5& 12& 60& 1.8& 3.4& $5\times 10^4$ & 10\\

\hline

1101+384(Mkn 421)& 0.03 &1& 0.06& 20& $3\times 10^5$  &  2.2& 5.3 &$4\times 10^4$ & -\\

1219+285(ON231)& 0.102& 0.7& 0.8& 14& $5\times 10^3$& 2& 3.9& $5\times
10^3$&-\\

1652+398(Mkn 501) & 0.03 &0.19& 0.32& 10& $1.1\times 10^5$  &  1.5& 3 &
750 & - \\

2155-304(PKS)& 0.117& 0.3 & 1 & 18 & $3.2\times 10^4$ & 2 & 4.85 &
$5\times 10^4$ & - \\

2200+420(BL Lac)& 0.07& 0.2& 1.5& 20& $10^3$& 1.9& 3.8& $2\times 10^5$ &-\\

\hline
\multicolumn{10}{l}{$^*$: see text} \\
\end{tabular}
\end{center}
\end{table*}

\newpage

\begin{table}
\begin{center}
\caption{Values of the estimated powers and luminosity for the sources
analyzed in the present work (see text for definitions).}

\begin{tabular}{lccccc}
\hline
\hline
Source & $L_{\rm disk}$ & $L_{\rm jet}$& $P_{\rm e}$ & $P_{B}$ & $P_{\rm jet}^*$\\

name & $10^{46}$ erg s$^{-1}$ & $10^{46}$ erg s$^{-1}$ & $10^{46}$ erg
s$^{-1}$ & $10^{46}$ erg s$^{-1}$ & $10^{46}$ erg s$^{-1}$\\ \hline
0208-512& 0.5& 0.7& 0.5& 0.12&50.8\\

0521-365& 0.024$^{**}$& 1.0$^{\rm a}$& 0.7& 0.05&5.4\\

0537-441& 2.0$^{**}$& 2.18& 0.6& 0.8&14.6\\

0836+710& 10.0& 17.3& 2.7& 2.7&500.0\\

1253-055& 0.2& 6.0& 0.2& 0.04&90.0\\

1510-089& 0.5& 0.12& 0.2& 0.6&50.0\\

1641+399& 3.7$^{**}$& 1.1& 0.15& 0.8&12.0\\

2223-052& 8.5$^{**}$& 1.2& 0.17& 8.2&24.7\\

2230-014& 2.0& 1.0& 0.2& 0.2&220.0\\

2243-123& 2.0& 0.4& 0.1& 0.83&12.0\\

2251+158& 6& 45.7& 1.0& 0.4&290.0\\

\hline

1101+384& $<5\times 10^{-4**}$& $5\times 10^{-4}$& 0.05& 0.001&42\\

1219+285& $<5\times 10^{-4**}$& $1\times 10^{-2}$& 0.004& 0.003 & 0.74\\

1652+398& $<2\times 10^{-4**}$& $5\times 10^{-3}$& 0.004& $1.5\times 10^{-5}$&0.014\\

2155-304& $<$0.017 & $2.5\times 10^{-4}$& 0.02& 0.002&2.5\\

2200+420& $2\times 10^{-4**}$& $2.3\times 10^{-2}$& 0.01& $4.5\times 10^{-4}$&1.4\\

\hline
\multicolumn{6}{l}{$^*$: calculated for $n_p=n_e$; $^{**}$: from CPG97
assuming $L_{disk}=10\times L_{BLR}$}\\
\multicolumn{6}{l}{$^{\rm a}$: calculated for $\Gamma =10$, see text}\\
\end{tabular}
\end{center}
\end{table}

\newpage

\begin{figure}
\centerline{\plotone{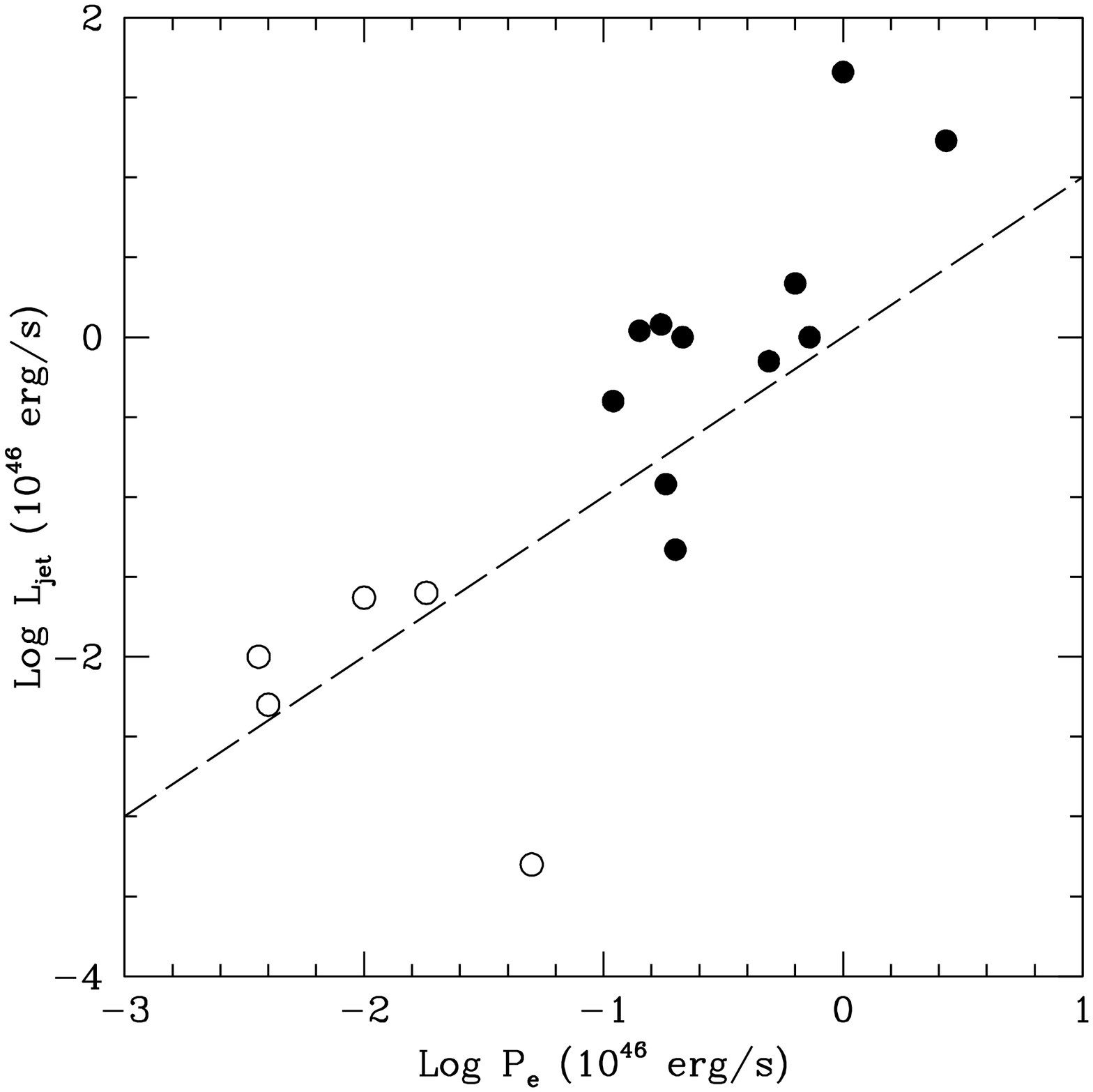}}
\end{figure}

\begin{figure}
\centerline{\plotone{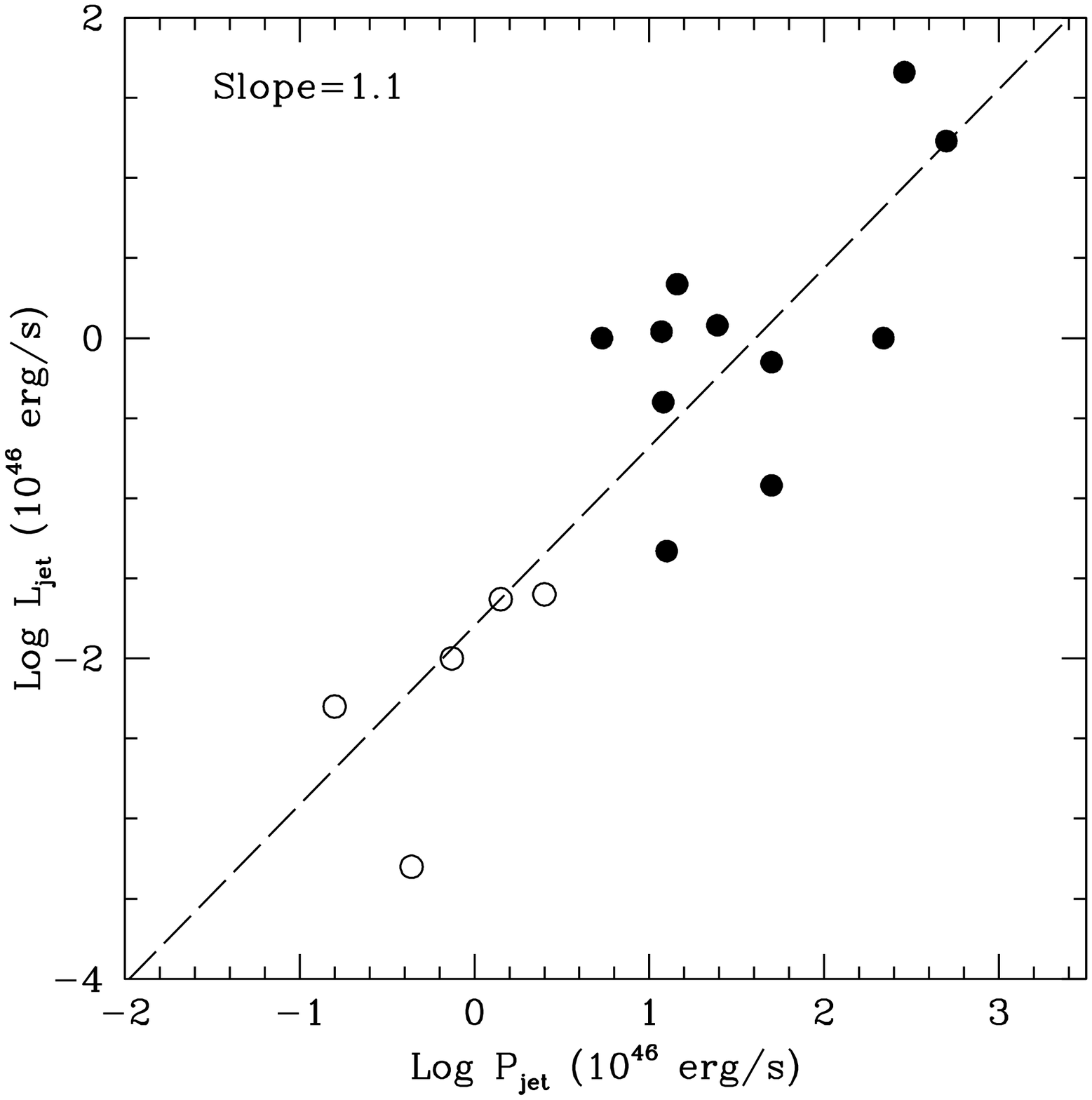}}
\end{figure}

\begin{figure}
\centerline{\plotone{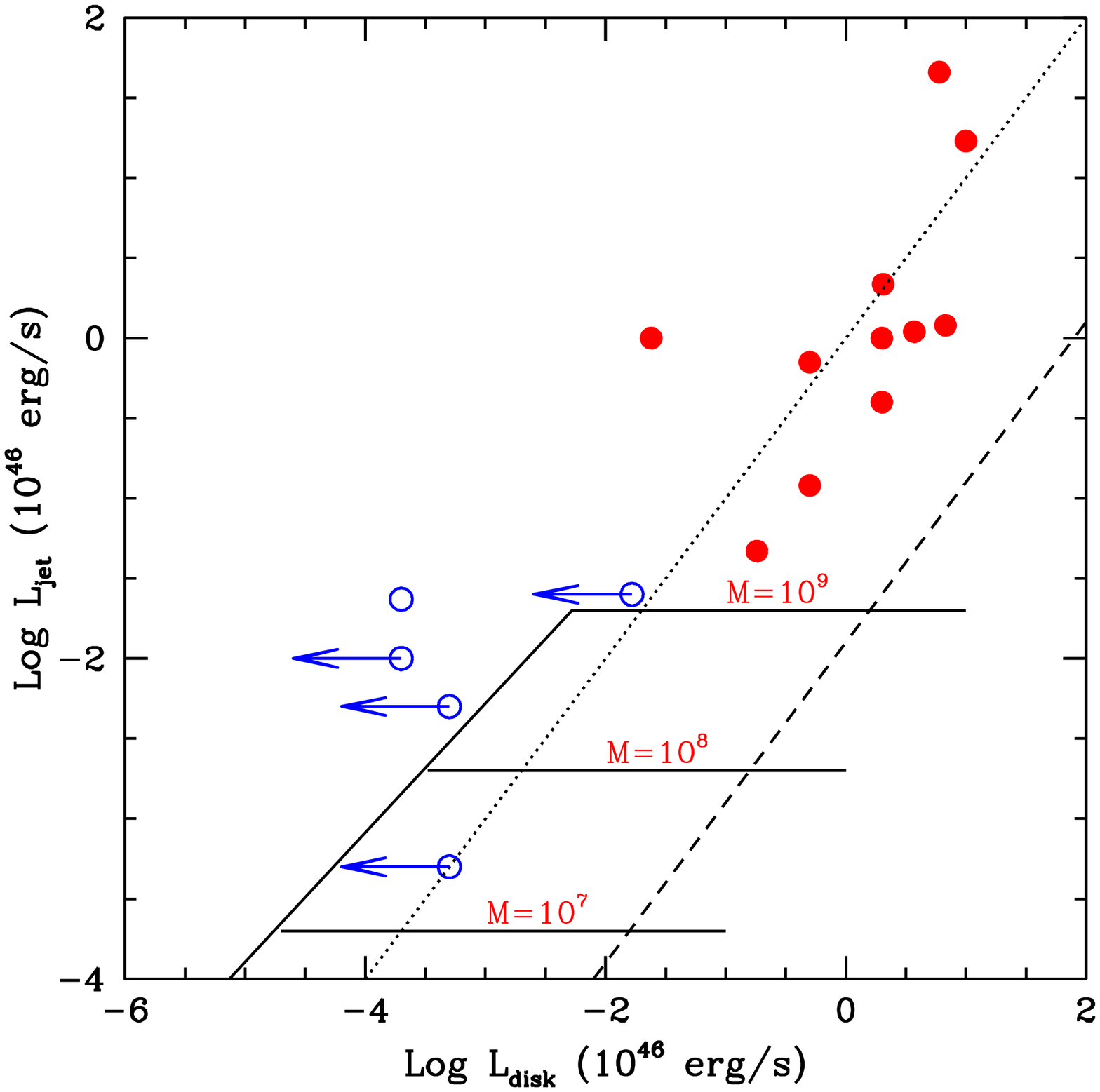}}
\end{figure}

\end{document}